\documentstyle[12pt]{article}
\begin{document}

%\begin{flushright}{October 31, 2005}
%\end{flushright}
%\vskip 0.5 truecm

\begin{center}
{\Large{\bf Quantum anomaly and geometric phase; their basic differences}}
\end{center}
\vskip .5 truecm
\centerline{\bf Kazuo Fujikawa }
\vskip .4 truecm
\centerline {\it Institute of Quantum Science, College of 
Science and Technology}
\centerline {\it Nihon University, Chiyoda-ku, Tokyo 101-8308, 
Japan}
\vskip 0.5 truecm

\makeatletter
\@addtoreset{equation}{section}
\def\theequation{\thesection.\arabic{equation}}
\makeatother

\begin{abstract}
It is sometimes stated in the literature that the quantum 
anomaly is regarded as an example of the geometric phase. 
Though there is some superficial similarity between these two 
notions, we here show that the differences bewteen these two 
notions are more profound and fundamental. As an explicit 
example, we analyze in detail a  quantum mechanical model 
proposed by M. Stone, which is supposed to show the above 
connection. We show that the geometric term in the model, which
is topologically trivial for any finite time interval $T$, 
corresponds to the so-called ``normal naive term'' in field 
theory and has nothing to do with the anomaly-induced 
Wess-Zumino term. In the fundamental level, the difference 
between the two notions is stated as follows: The topology of 
gauge fields leads to level crossing in the fermionic sector
 in the case of chiral anomaly and the {\em failure} of the 
adiabatic approximation is essential in the analysis, whereas 
the (potential) level crossing in the matter sector leads to the 
topology of the Berry phase only when the precise adiabatic 
approximation holds.
\end{abstract}

%\large

\section{Introduction}

In quantum field theory the quantum anomaly plays an important
role to test if a specific classical symmetry in 
question is really preserved in quantum 
theory~\cite{adler, jackiw, bertlmann, fuji-suzuki}. The quantum
anomaly also predicts some novel phenomena which are not 
expected by a classical consideration, for example, the baryon 
number violation in the Weinberg-Salam theory~\cite{'t hooft2}. 
In some special 
cases of chiral anomaly, one can summarize the effects of the
quantum  anomaly in the form of an extra Wess-Zumino 
term~\cite{wess} which is added to the starting Lagrangian. 

On the other hand, it has been recognized that one 
obtains phase factors in the adiabatic treatment (
such as in the Born-Oppenheimer approximation) of 
the Schr\"{o}dinger equation which depends on  
slowly varying background variables~\cite{pancharatnam}
-\cite{berry2}. These phases are called 
``geometric phases'', and they are generally associated with
level crossing. Although the manner of obtaining 
geometric phases is quite different from that of quantum 
anomalies, it is sometimes stated in the literature that the 
chiral anomaly is regarded as a kind of geometric 
phase~\cite{stone, aitchison}.

The notion of the geometric phase itself does not appear to be 
sharply defined at this moment. In the influential book by 
Shapere and Wilczek~\cite{shapere}, various phase factors in 
physics which exhibit topological properties are discussed 
together as geometric phases.
It is important to synthesize various phenomena and notions 
into a unifying notion, but it is the opinion of the present 
author that this broad use of the scientific terminology could 
lead to confusions and mis-understandings in view of the wide 
use of geometric phases in various fields in physics today. 
This broad use of the terminology is closely related to the 
broad use of the terminology of ``adiabatic approximation''.
The practical Born-Oppenheimer approximation, which provides a 
typical adiabatic approximation in physics, contains two 
quite different time scales but the slower time scale $T$ 
measured in units of the time scale of the faster system is 
{\em finite}. In such a practical Born-Oppenheimer 
approximation, it is shown that the commonly referred Berry's 
phase, which is purely dynamical without any approximation,  
is topologically trivial and no monopole-like singularity at 
the level crossing point~\cite{fujikawa2}. The notion such as  
holonomy is valid for the level crossing problem only in the 
precise adiabatic limit with $T\rightarrow\infty$~\cite{simon}. 

The above properties of the geometric phase become quite clear 
in a recent attempt to formulate the geometric phase in the 
second quantized formulation~\cite{fujikawa2}. This 
approach  works in both of path integral and operator 
formulations, and the analysis of geometric phases is reduced to 
a simple diagonalization of the Hamiltonian. The hidden local
gauge symmetry, which arises from the fact that the choice of 
basis vectors in the functional space is arbitrary in field 
theory, replaces the notions of parallel transport and 
holonomy~\cite{fujikawa3}. By carefully 
diagonalizing the geometric term in the infinitesimal 
neighborhood of level crossing, it is  shown that the 
topological property of the geometric phase is 
trivial in the practical Born-Oppenheimer approximation, where 
the period $T$ of the slower system is finite, and 
thus no monopole-like singularity, as already stated above. 
This approximate topology in the geometric phase is 
quite different from the exact topology associated with gauge 
fields such as in the familiar Aharonov-Bohm 
effect~\cite{aharonov}. We thus 
become somewhat suspicious about the claim on the equivalence of
 quite distinct notions such as quantum anomaly and geometric 
phase. The purpose of the present paper is to show that these 
two notions, namely, quantum anomaly and geometric phase, may 
have some superficial similarity to each other, but the 
differences in these two notions are more profound and 
fundamental.

In the literature, the paper by M. Stone~\cite{stone} is often 
quoted as an evidence of the equivalence of the quantum anomaly 
and the geometric phase. We thus  explain the crucial 
differences between the geometric phase and the quantum 
anomaly by taking the model by Stone~\footnote{It should 
however be emphasized that we are not criticizing the analysis 
of the geometric phase itself in the model by 
Stone, which is essentially identical to the 
simplest example in~\cite{berry}.} as a concrete example, 
though our analysis is valid for the more general model 
summarized in Appendix.
We first analyze the problem from the point of view of several
characteristic properties of the chiral 
anomaly~\footnote{To make 
our analysis definite, we define the quantum anomaly as
the even-dimensional chiral anomaly and the geometric phase as 
the phase associated with general level crossing which is 
summarized in Appendix. }, such as the failure of the
naive manipulation and the failure of the decoupling theorem,
on the basis of the explicit model in~\cite{stone} and a
corresponding field theoretical model which contains a true 
anomaly in Sections 2 and 3.  We 
show that the interpretation of the geometric term in the model 
in~\cite{stone} as the Wess-Zumino term, namely, a manifestation
 of quantum anomaly is untenable even in the precise adiabatic 
approximation. We then analyze the problem from the point of 
view of two key concepts involved in both of the chiral anomaly 
and the geometric phase, namely, level crossing and topology.
By a careful examination of the statements made in the paper by
Nelson and Alvarez-Gaume~\cite{nelson}, we explain in Section
4 that the chiral anomaly and the geometric phase are completely
 different in the fundamental level.  

\section{Quantum mechanical model; geometric phase}

We first recapitulate the model due to
M. Stone\cite{stone}. The model  starts with the Hamiltonian
\begin{eqnarray}
H=\frac{\vec{L}^{2}}{2I}-\mu{\bf n}(t)\cdot\vec{\sigma}
\end{eqnarray}
where ${\bf n}(t)$ is a unit vector specifying the direction of 
the ``magnetic field'' acting on the spin represented by the 
Pauli matrix $\vec{\sigma}$, and $\vec{L}$ generates the 
rotation of ${\bf n}(t)$.  
We analyze the mathematical aspects of the model (2.1) in this 
paper without asking the possible physical meaning of the 
specific model, which is explained in~\cite{stone}.
Partly referring to the second quantization, one can write the 
above Hamiltonian as
\begin{eqnarray}
H(t)=\frac{\vec{L}^{2}}{2I}-\psi^{\dagger}\mu{\bf n}(t)\cdot
\vec{\sigma}\psi
\end{eqnarray}
where the field $\psi$ stands for the two-component spinor.

One may then write an evolution operator in the formal path
integral representation~\cite{stone}
\begin{eqnarray}
&&\langle f|\exp[-\frac{i}{\hbar}\int_{0}^{T}H dt]|i\rangle
\nonumber\\
&&=\int{\cal D}\vec{n}
{\cal D}\psi^{\dagger}{\cal D}\psi\delta(\vec{n}^{2}-1)\nonumber
\\
&&\times\exp\{\frac{i}{\hbar}\int_{0}^{T}dt
[\frac{\dot{\vec{n}}^{2}}{2I}
+\psi^{\dagger}\frac{\hbar}{i}\partial_{t}\psi
+\psi^{\dagger}\mu\vec{n}(t)\cdot\vec{\sigma}\psi] \}
\end{eqnarray}
or in the Euclidean formulation ($t\rightarrow -i\tau$), we have 
\begin{eqnarray}
&&\int{\cal D}\vec{n}
{\cal D}\psi^{\dagger}{\cal D}\psi\delta(\vec{n}^{2}-1)
\nonumber\\
&&\times\exp\{\frac{1}{\hbar}\int_{0}^{\beta}d\tau
[-\frac{\dot{\vec{n}}^{2}}{2I}
+\psi^{\dagger}\hbar\partial_{\tau}\psi
+\psi^{\dagger}\mu\vec{n}(\tau)\cdot\vec{\sigma}\psi] \}.
\end{eqnarray}
Following Ref.~\cite{stone}, we take this path integral as our
starting point. 

This path integral is rewritten as 
\begin{eqnarray}
&&\int{\cal D}\vec{n}
{\cal D}{\psi^{\prime}}^{\dagger}{\cal D}\psi^{\prime}
\delta(\vec{n}^{2}-1)\nonumber\\
&&\times\exp\{\frac{1}{\hbar}\int_{0}^{\beta}d\tau
[-\frac{\dot{\vec{n}}^{2}}{2I}
+{\psi^{\prime}}^{\dagger}(\hbar\partial_{\tau}
+\mu\sigma_{3}+U(\vec{n}(\tau))^{\dagger}
\hbar\partial_{\tau}U(\vec{n}(\tau)))\psi^{\prime}] \}
\nonumber\\
&&=\int{\cal D}\vec{n}
{\cal D}{\psi}^{\dagger}{\cal D}\psi
\delta(\vec{n}^{2}-1)\nonumber\\
&&\times\exp\{\frac{1}{\hbar}\int_{0}^{\beta}d\tau
[-\frac{\dot{\vec{n}}^{2}}{2I}
+{\psi}^{\dagger}(\hbar\partial_{\tau}
+\mu\sigma_{3}+U(\vec{n}(\tau))^{\dagger}
\hbar\partial_{\tau}U(\vec{n}(\tau)))\psi]\}
\nonumber\\
\end{eqnarray}
when one performs a time-dependent unitary transformation
(or a gauge transformation)
\begin{eqnarray}
\psi(\tau)&=& U(\vec{n}(\tau))\psi^{\prime}(\tau),\nonumber\\
\psi^{\dagger}(\tau)&=&{\psi^{\prime}}^{\dagger}(\tau) 
U^{\dagger}(\vec{n}(\tau))
\end{eqnarray}
with
\begin{eqnarray}
U(\vec{n}(\tau))^{\dagger}\vec{n}(\tau)\vec{\sigma}
U(\vec{n}(\tau))
=|\vec{n}|\sigma_{3}.
\end{eqnarray}
The last relation in (2.5) means that the naming of 
integration variables is arbitrary in the path integral.
An explicit form of the unitary transformation is given by
defining
\begin{eqnarray}
&&v_{+}(\vec{n})=\left(\begin{array}{c}
            \cos\frac{\theta}{2}e^{-i\varphi}\\
            \sin\frac{\theta}{2}
            \end{array}\right), \nonumber\\ 
&&v_{-}(\vec{n})=\left(\begin{array}{c}
            \sin\frac{\theta}{2}e^{-i\varphi}\\
            -\cos\frac{\theta}{2}
            \end{array}\right)
\end{eqnarray}
in terms of the polar coordinates, 
$n_{1} = |\vec{n}|\sin\theta\cos\varphi$,
$n_{2} = |\vec{n}|\sin\theta\sin\varphi$, 
$n_{3} = |\vec{n}|\cos\theta$. Note that these eigenfunctions,
which satisfy
\begin{eqnarray}
\mu\vec{n}(\tau)\vec{\sigma}v_{\pm}(\vec{n})=\pm\mu|\vec{n}|
 v_{\pm}(\vec{n}),
\end{eqnarray}
are singular at the origin $\mu\vec{n}=0$ and also contain 
spurious singularities at north and south poles~\footnote{In the context of level crossing, it is natural to consider the combination $\mu\vec{n}$ by allowing the  possible time 
dependence of $\mu(t)$. If the variable $\mu(t)\vec{n}$ moves toward the origin during a cyclic motion, it implies that the two levels appraoch the level crossing point. }. In our choice 
of the phase convention, we have 
$v_{\pm}(\vec{n}(0))= v_{\pm}(\vec{n}(\beta))$ if 
$\vec{n}(0)=\vec{n}(\beta)$; it has been explained in detail 
elsewhere~\cite{fujikawa3} and later in the appendix that
 the choice of the time-dependent phases of these eigenfunctions
is arbitrary due to the hidden local gauge symmetry.  
Then $U(\vec{n}(\tau))$ is given by a $2\times2$ matrix
\begin{eqnarray}
U(\vec{n}(\tau))=\left(\begin{array}{cc}
             v_{+}(\vec{n})\  v_{-}(\vec{n})
               \end{array}\right).
\end{eqnarray} 
This unitary transformation keeps the path integral measure
invariant
\begin{eqnarray}
{\cal D}\psi^{\dagger}{\cal D}\psi=
{\cal D}{\psi^{\prime}}^{\dagger}{\cal D}\psi^{\prime}
\end{eqnarray}
without giving a non-trivial Jacobian for the present 
two-component  problem (2.6), as long as 
$U(\vec{n}(\tau))$ is  not singular. The matrix 
$U(\vec{n}(\tau))$ 
becomes singular at the level crossing point which takes place 
at $\mu\vec{n}=0$ in the present case. (In terms of the polar 
coordinates, $U(\vec{n}(\tau))$ at the north or south pole 
exhibits spurious singularity.) The treatment in the 
infinitesimal neighborhood of the singularity is discussed later.

If one defines (in Euclidean metric, but the result is valid for 
Minkowski metric also)
\begin{eqnarray} 
v^{\dagger}_{m}(\vec{n})i\frac{\partial}{\partial \tau}
v_{n}(\vec{n})
=A_{mn}^{k}(\vec{n})\dot{n}_{k} 
\end{eqnarray}
where $m$ and $n$ run over $\pm$,
we have
\begin{eqnarray}
A_{++}^{k}(\vec{n})\dot{n}_{k}
&=&\frac{(1+\cos\theta)}{2}\dot{\varphi},
\nonumber\\
A_{+-}^{k}(\vec{n})\dot{n}_{k}
&=&\frac{\sin\theta}{2}\dot{\varphi}+\frac{i}{2}\dot{\theta}
=(A_{-+}^{k}(\vec{n})\dot{n}_{k})^{\star}
,\nonumber\\
A_{--}^{k}(\vec{n})\dot{n}_{k}
&=&\frac{1-\cos\theta}{2}\dot{\varphi}. 
\end{eqnarray}
Note that we have
\begin{eqnarray}
{\rm Tr}[v^{\dagger}_{m}(\vec{n})i\frac{\partial}{\partial \tau}
v_{n}(\vec{n})]=\dot{\varphi}.
\end{eqnarray}

The above relation (2.5) implies the equivalence of two 
Lagrangians
\begin{eqnarray}
 L =-\frac{\dot{\vec{n}}^{2}}{2I}
+\psi^{\dagger}\hbar\partial_{\tau}\psi
+\psi^{\dagger}\mu\vec{n}(\tau)\cdot\vec{\sigma}\psi
\end{eqnarray}
and 
\begin{eqnarray}
L^{\prime}=-\frac{\dot{\vec{n}}^{2}}{2I}
+{\psi}^{\dagger}(\hbar\partial_{\tau}
+\mu\sigma_{3}+U(\vec{n}(\tau))^{\dagger}
\hbar\partial_{\tau}U(\vec{n}(\tau)))\psi. 
\end{eqnarray}
The fermionic part of the starting Hamiltonian (2.2) is thus 
equivalent 
to~\footnote{The Legendre transformation from the Lagrangian to
the total Hamiltonian is involved in the presence of the 
derivative coupling as in the present example (2.16). Thus our 
fermionic  
Hamiltonian (2.17) is valid only when the variable $\vec{L}$ or 
$n(t)$ is treated as a background c-number. This limitation, 
however, does not influence our analysis of the possible 
connection of the geometric term with the Wess-Zumino term. The 
analysis of geometric term is generally performed in this 
simplified situation~\cite{berry2}. 
The second quantized path integral approach to the geometric 
term~\cite{fujikawa2} is more flexible for the treatment of  
more general situations.} (by going back to Minkowski metric)
\begin{eqnarray}
H_{\rm fermion}(t)=-{\psi}^{\dagger}
[\mu\sigma_{3}+ U(\vec{n}(t))^{\dagger}
\frac{\hbar}{i}\partial_{t}U(\vec{n}(t))|_{|\vec{n}|=1}]
\psi
\end{eqnarray}
or in the original notation of (2.1)
\begin{eqnarray}
H_{\rm fermion}=-\mu\sigma_{3}
-U(\vec{n}(t))^{\dagger}
\frac{\hbar}{i}\partial_{t}U(\vec{n}(t))|_{|\vec{n}|=1}.
\end{eqnarray}
The last term in (2.18), which may be  understood as a pure 
gauge term, is generally called  as "geometric term" for the 
historical reason. The survival of this geometric term in
the limit of the large $\mu$ limit was interpreted in 
Ref.~\cite{stone} as an evidence of the failure of the 
decoupling theorem. The failure of the decoupling theorem in 
the context of quantum anomaly is however more involved, as 
will be explained later.  
This Hamiltonian (2.18), which is exact, carries all the 
information about the geometric phases as we show below; this 
means that the geometric phases are purely {\em dynamical}. 

If one is interested in the lower energy state of 
the Hamiltonian (2.18), one
has an approximate Hamiltonian
\begin{eqnarray}
H_{ad}&\simeq& -\mu
-(U(\vec{n}(t))^{\dagger}
\frac{\hbar}{i}\partial_{t}U(\vec{n}(t))|_{|\vec{n}|=1})_{++}
\nonumber\\
&=& -\mu + \hbar
\frac{(1+\cos\theta)}{2}\dot{\varphi}
\end{eqnarray}
by noting (2.13).
If $\mu$ is sufficiently large, to be precise
for
\begin{eqnarray}
2\mu T\gg 2\pi\hbar,
\end{eqnarray}
one may neglect the off-diagonal part in the geometric term in
(2.18), and 
this Hamiltonian $H_{ad}$ provides a good adiabatic 
approximation to the 
full Hamiltonian. Here $T$ is the period of the slower dynamical
system $\vec{n}(t)$ and $2\pi\hbar$ stands for the magnitude of 
the geometric term times $T$. We emphasize that the adiabatic 
approximation in the present context corresponds to throwing 
away the off-diagonal part in the geometric term, 
namely, throwing away {\em a part of the Hamiltonian}. The 
geometric term in (2.19) is reminiscent of a magnetic monopole 
located in the parameter space at the level 
crossing point $\mu\vec{n}=0$. The fermionic 
Hamiltonian (2.19) thus gives rise to the dynamical phase
\begin{eqnarray}
&&\exp\{-\frac{i}{\hbar}\int_{0}^{T}dt[-\mu + \hbar
\frac{(1+\cos\theta)}{2}\dot{\varphi}]\}\nonumber\\
&&=\exp\{\frac{i}{\hbar}\mu T - i \oint 
\frac{(1+\cos\theta)}{2} d\varphi]\}
\end{eqnarray}
for a cyclic motion of the slower sytem, and the second term
gives rise to the familiar Berry's phase~\cite{stone, berry}.

The last geometric term in (2.18) has an {\em approximate} 
topological 
property around the level crossing point in the practical 
Born-Oppenheimer approximation where the period of the slower 
dynamical system $T$ is finite. This fact is understood as 
follows: For sufficiently close to the 
level crossing point, $\mu\sim 0$ but $\mu\neq0$, one has 
$\mu T \ll 2\pi\hbar$ instead of (2.20). 
One may then perform a further unitary transformation of the 
fermionic  variable~\cite{fujikawa2}
\begin{eqnarray}
\psi^{\prime}(t)&=& U(\theta(t))\psi^{\prime\prime}(t),
\nonumber\\
{\psi^{\prime}(t)}^{\dagger}&=&
{\psi^{\prime\prime}}^{\dagger}(t) 
U^{\dagger}(\theta(t))
\end{eqnarray}
with
\begin{eqnarray}
U(\theta(t))=\left(\begin{array}{cc}
            \cos\frac{\theta}{2}&-\sin\frac{\theta}{2}\\
            \sin\frac{\theta}{2}&\cos\frac{\theta}{2}
            \end{array}\right)
\end{eqnarray}
in addition to (2.6).
The Hamiltonian (2.17) is thus equivalent to (by repeating the 
path integral analysis) 
\begin{eqnarray}
H_{\rm fermion}(t)
&=&-{\psi}^{\dagger}
[\mu U(\theta(t))^{\dagger}
\sigma_{3}U(\theta(t))\nonumber\\
&&+ (U(\theta(t))U(\vec{n}(t)))^{\dagger}
\frac{\hbar}{i}\partial_{t}(U(\vec{n}(t))U(\theta(t)))|_{|\vec{n}|=1}]
\psi\nonumber\\
&=&-{\psi}^{\dagger}[
\mu U(\theta(t))^{\dagger}
\sigma_{3}U(\theta(t))-
\hbar\left(\begin{array}{cc}
            \dot{\varphi}&0\\
            0&0
            \end{array}\right)]\psi
\nonumber\\
&\simeq&+ {\psi}^{\dagger}
\hbar\left(\begin{array}{cc}
            \dot{\varphi}&0\\
            0&0
            \end{array}\right)\psi
\end{eqnarray}
for $\mu \sim 0$, or in the original 
notation
\begin{eqnarray}
H_{\rm fermion}\simeq \hbar\left(\begin{array}{cc}
            \dot{\varphi}&0\\
            0&0
            \end{array}\right).
\end{eqnarray} 
The geometric phase thus either vanishes or becomes trivial
\begin{eqnarray}
\exp\{-i\int_{0}^{T} \dot{\varphi}dt\}=\exp\{-2i\pi\}=1
\end{eqnarray}
in the infinitesimal neighborhood of level crossing.
The geometric term  is thus topologically (i.e., under the 
continuous variation of the parameter $\mu$) trivial for any 
finite $T$.
At the level crossing point, $\mu \sim 0$,
the conventional energy becomes 
degenerate but the degeneracy is lifted when one diagonalizes 
the geometric term. It is important that the 
additional transformation (2.23) depends on the variable 
$\theta$ only and preserves (2.14).

Though the geometric phase is topologically trivial in a 
precise sense, it 
is still interesting that the geometric phase is approximately 
topological. This approximate topological property (of a pure 
gauge term) is traced to the fact that the eigenfunctions in 
(2.9) are singular on top of the level crossing, {\em i.e.}, the 
gauge transformation (2.6) is singular, though the 
singular behavior is avoided in the sense of the trivial phase 
as in (2.25) by defining a suitable basis set in the 
neighborhood of the singularity by a further unitary 
transformation. (The above relation (2.24) also shows 
that if the time variation of $\vec{n}(t)$ is faster than the 
fermionic variables even for $\mu$ which is not small, 
the geometric term dominates the $\mu\sigma_{3}$ term,
and the geometric term becomes topologically trivial. This is 
another idication that the geometric term is not quite 
topological, and this observation becomes relevant when one 
compares the geometric phase with the chiral anomaly. )

The geometric 
term corresponds to the {\em normal term} not an anomalous 
term in field theory, as we explain in Section 3. The geometric term in
 the present model has nothing to do with the Wess-Zumino term 
as we understand it in field theory which is a result of the 
symmetry breaking by quantum effects. To be more precise, (2.5)
 shows  that 
\begin{eqnarray}
&&{\rm det}[\hbar\partial_{\tau}
+\mu\vec{n}(\tau)\cdot\vec{\sigma}]\nonumber\\
&&={\rm det}[U(\vec{n}(\tau))^{\dagger}\hbar\partial_{\tau}
+\mu\vec{n}(\tau)\cdot\vec{\sigma}\}U(\vec{n}(\tau))]\nonumber\\
&&={\rm det}[\hbar\partial_{\tau}
+\mu\sigma_{3}+U(\vec{n}(\tau))^{\dagger}
\hbar\partial_{\tau}U(\vec{n}(\tau))].
\end{eqnarray}
The ordinary Wess-Zumino term would manifest itself as an 
{\em extra} phase factor on the right-hand side of this relation
 (see, for example, (3.25)), but no such an extra phase in the 
present example. 

This analysis of the Wess-Zumino term becomes more transparent 
if one considers $H=\frac{\vec{L}^{2}}{2I}
-\mu{\bf n}(t)\cdot\vec{\sigma} +\mu_{0}$ instead of (2.1), or
\begin{eqnarray}
H=\frac{\vec{L}^{2}}{2I}
-\psi^{\dagger}[\mu{\bf n}(t)\cdot\vec{\sigma}-\mu_{0}]\psi
\end{eqnarray}
instead of (2.2) with a positive constant $\mu_{0}$ which 
satisfies
\begin{eqnarray}
\mu_{0}>\mu
\end{eqnarray}
by noting that the absolute value of the energy is not fixed 
in the present quantum mechanical model. This choice incidentally
defines a Euclidean theory more precisely, as the Hamiltonian 
becomes positive definite.
Then the equivalent Hamiltonian (by treating $\vec{n}(t)$ as a
background variable) in (2.17) is replaced by 
\begin{eqnarray}
H_{\rm fermion}(t)=-{\psi}^{\dagger}
[\mu\sigma_{3}- \mu_{0}+ U(\vec{n}(t))^{\dagger}
\frac{\hbar}{i}\partial_{t}U(\vec{n}(t))|_{|\vec{n}|=1}]
\psi
\end{eqnarray}
or in the original notation of (2.1)
\begin{eqnarray}
H_{\rm fermion}=-\mu\sigma_{3} + \mu_{0}
-U(\vec{n}(t))^{\dagger}
\frac{\hbar}{i}\partial_{t}U(\vec{n}(t))|_{|\vec{n}|=1}.
\end{eqnarray}
The adiabatic approximation for the lower energy state 
$|+\rangle$ is then given by 
\begin{eqnarray}
H_{ad}&\simeq& -\mu + \mu_{0}
-(U(\vec{n}(t))^{\dagger}
\frac{\hbar}{i}\partial_{t}U(\vec{n}(t))|_{|\vec{n}|=1})_{++}
\nonumber\\
&=& -\mu + \mu_{0} + \hbar
\frac{(1+\cos\theta)}{2}\dot{\varphi}
\end{eqnarray}
and the dynamical phase for the fermionic part is given by
\begin{eqnarray}
&&\exp\{-\frac{i}{\hbar}\int_{0}^{T}dt[-\mu + \mu_{0} + \hbar
\frac{(1+\cos\theta)}{2}\dot{\varphi}]\}\nonumber\\
&&=\exp\{-\frac{i}{\hbar}(\mu_{0} - \mu)T - i \oint 
\frac{(1+\cos\theta)}{2} d\varphi]\}.
\end{eqnarray}
We thus obtain the same geometric phase independently of 
$\mu_{0}$. The almost topological property of the geometric 
phase arises from the crossing of two levels 
\begin{eqnarray}
\mu_{0}\pm \mu|\vec{n}(t)|>0
\end{eqnarray}
at $\mu|\vec{n}(t)|=0$; the crossing of {\em positive} and 
{\em negative} levels at $\mu|\vec{n}(t)|=0$, which is realized 
when one sets $\mu_{0}=0$, is not essential for the geometric
phase. The fact that we can include an arbitrary mass parameter 
$\mu_{0}$ shows that the basic symmetry in the present model is 
{\em vector-like}
which contains no anomaly, to be consistent with the absence of 
the non-trivial Jacobian. This may be compared to (3.7).

For the present case (2.28) also, we have a naive relation
\begin{eqnarray}
&&{\rm det}[\hbar\partial_{\tau}
+\mu\vec{n}(\tau)\cdot\vec{\sigma}-\mu_{0}]\nonumber\\
&&={\rm det}[U(\vec{n}(\tau))^{\dagger}\{\hbar\partial_{\tau}
+\mu\vec{n}(\tau)\cdot\vec{\sigma}-\mu_{0}\}U(\vec{n}(\tau))]
\nonumber\\
&&={\rm det}[\hbar\partial_{\tau}
+\mu\sigma_{3}- \mu_{0} +U(\vec{n}(\tau))^{\dagger}
\hbar\partial_{\tau}U(\vec{n}(\tau))]
\end{eqnarray}
without any extra phase factor which would correspond to the 
Wess-Zumino term. We also have for $H_{\rm fermion}$ in (2.30)
\begin{eqnarray}
\langle 0|\exp\{-\frac{1}{\hbar}\int_{0}^{\beta} 
H_{\rm fermion}(\tau)d\tau\}|0\rangle
=1
\end{eqnarray}
for the fermionic vacuum $|0\rangle$ in the second quantized 
sense defined by 
\begin{eqnarray}
\psi_{+}|0\rangle=\psi_{-}|0\rangle=0
\end{eqnarray}
in the adiabatic picture where one can approximately diagonalize
 the fermionic Hamiltonian by treating the variable $\vec{n}(t)$
 as a background c-number. Note
that the energies of the fermionic states are positive definite with vanishing vacuum energy in the adiabatic picture. 
The important point here is that we do not have any extra phase
in (2.35), and we do not have any contribution from the fermionic
part of the Hamiltonian for the evolution operator (2.36). This is consistent with the general relation
\begin{eqnarray}
&&{\rm det}[\hbar\partial_{\tau}
+\mu\sigma_{3}- \mu_{0} +U(\vec{n}(\tau))^{\dagger}
\hbar\partial_{\tau}U(\vec{n}(\tau))]
\nonumber\\
&&={\rm Str}\{\exp{[-(1/\hbar)\int_{0}^{\beta}
H_{\rm fermion}(\tau)d\tau]}\}\nonumber\\
&&\sim \exp\{-\frac{1}{\hbar}\langle 0|H_{\rm fermion}
|0\rangle \beta\}=1
\end{eqnarray}
for $\beta\rightarrow \infty$ with fixed large $\mu$ 
and $\mu_{0}$ with $\mu_{0}>\mu$ such that the vacuum with 
vanishing energy is isolated. When one defines the functional 
determinant with periodic boundary conditions, the 
determinant gives a supertrace. 
If one should have a Wess-Zumino term, the both-hand sides of 
this relation (2.38) would have an extra non-trivial phase 
factor relative to ${\rm det}[\hbar\partial_{\tau}
+\mu\vec{n}(\tau)\cdot\vec{\sigma}-\mu_{0}]$. See eq.(3.25). 

Instead of (2.38), one might prefer to consider (2.27) for 
$\beta \rightarrow \rm large$
\begin{eqnarray}
&&{\rm det}[\hbar\partial_{\tau}
+\mu\sigma_{3}+U(\vec{n}(\tau))^{\dagger}
\hbar\partial_{\tau}U(\vec{n}(\tau))]
\nonumber\\
&&\sim \exp\{-\frac{1}{\hbar}\int_{0}^{\beta}d\tau
\langle +|H_{\rm fermion}
|+\rangle\}\nonumber\\
&&=\exp\{\frac{\mu\beta}{\hbar} - i \oint 
\frac{(1+\cos\theta)}{2} d\varphi\}
\end{eqnarray}
with the fermionic Hamiltonian (2.17), for which one-fermion
state with up-spin gives the energy lower than the 
vacuum~\cite{stone}. It 
thus appears that one obtains the geometric term from the 
fermionic functional determinant in the leading term. This 
relation (2.39) is however ill-defined for 
$\beta\rightarrow\infty$ for which the geometric (adiabatic)
phase is best defined~\cite{simon}. Also, the vacuum and 
the state $|+-\rangle$ are degenerate in this case
\begin{eqnarray}
&&\exp\{-\int_{0}^{\beta}d\tau\langle+-|H_{\rm fermion}|+-\rangle
\}\nonumber\\
&&=\exp\{-\int_{0}^{\beta}d\tau\langle+|H_{\rm fermion}|+\rangle
-\int_{0}^{\beta}d\tau\langle-|H_{\rm fermion}|-\rangle\}
\nonumber\\
&&=\exp\{\frac{\mu\beta}{\hbar} - i \oint 
\frac{(1+\cos\theta)}{2} d\varphi-\frac{\mu\beta}{\hbar} 
- i \oint \frac{(1-\cos\theta)}{2} d\varphi\}\nonumber\\
&&=\exp\{- i \oint d\varphi\}=\exp\{- 2\pi i\}=1.
\end{eqnarray}

It should be noted that the geometric terms appear in the 
sub-leading terms in (2.38). In this respect, it is immaterial 
if the geometric terms appear in the leading term or in the 
sub-leading terms by varying the parameter $\mu_{0}$. The 
crucial property is that the Wess-Zumino 
term, if it should exist in the present model, should appear
 multiplying {\em all} the terms, not only the leading term but 
also the sub-leading terms in both of (2.38) and (2.39) when 
one starts,
respectively, with the left-hand sides of (2.35) and (2.27). 
Obviously, no such a Wess-Zumino term in the present model. This may be compared to (3.26).
\\

\section{Field theoretical model; quantum anomaly}

A unitary transformation and induced terms which are analogous
to those discussed in the preceding section are realized by a 
field theoretical model defined by 
\begin{eqnarray}
{\cal L}&=&\bar{\psi}(x)[i\gamma^{\mu}(\partial_{\mu} 
-ieQA_{\mu})-
mU(\pi)]\psi(x)
\nonumber\\
&+&\frac{f^{2}_{\pi}}{16}{\rm Tr}\partial_{\mu}U(\pi)
\partial^{\mu}U(\pi)^{\dagger}
\end{eqnarray}
where
\begin{eqnarray}
U(\pi)=e^{2i(1/f_{\pi})\gamma_{5}\pi^{a}(x)T^{a}}
\end{eqnarray}
and
\begin{eqnarray}
&&\psi(x)=
\left(\begin{array}{c}
            p(x)\\
            n(x)
            \end{array}\right),\nonumber\\
&&Q=\left(\begin{array}{cc}
            1& 0\\
            0&0
            \end{array}\right), \ \ \ 
T^{a}=\frac{1}{2}\sigma^{a}.
\end{eqnarray}
In the present field theoretical model, we work in the Euclidean
metric with $g_{\mu\nu}=(-1,-1,-1,-1)$. In this model
$p(x)$ and $n(x)$ stand respectively for the idealized 
proton and 
neutron which are degenerate in mass, and $\pi^{a}(x)$ stand for
 the pion fields with
$\sigma^{a}$ standing for the Pauli matrix. $A_{\mu}(x)$ is the 
electromagnetic field. The above Lagrangian is invariant under
the electromagnetic gauge transformation and also invariant under
a global chiral $SU_{L}(2)\times SU_{R}(2)$ transformation which
 is weakly broken by the electromagnetic interaction. This chiral
symmetry becomes explicit by writing the above Lagrangian as
\begin{eqnarray}
{\cal L}&=&\bar{\psi}_{L}(x)i\gamma^{\mu}(\partial_{\mu} -ieQA_{\mu})\psi_{L}(x)+\bar{\psi}_{R}(x)i\gamma^{\mu}(\partial_{\mu} -ieQA_{\mu})\psi_{R}(x)\nonumber\\
&&-\bar{\psi}_{L}(x)me^{2i(1/f_{\pi})\pi^{a}(x)T^{a}}\psi_{R}(x)
-\bar{\psi}_{R}(x)
me^{-2i(1/f_{\pi})\pi^{a}(x)T^{a}}\psi_{L}(x)\nonumber\\
&&+\frac{f^{2}_{\pi}}{16}{\rm Tr}\partial_{\mu}U(\pi)
\partial^{\mu}U(\pi)^{\dagger}
\end{eqnarray}
where 
\begin{eqnarray}
\psi_{L,R}(x)=(\frac{1\mp\gamma_{5}}{2})\psi(x).
\end{eqnarray}
Under the global chiral transformation with global parameters
$\chi^{a}T^{a}$,
\begin{eqnarray}
\psi_{L}(x)&=& e^{-i\chi^{a}T^{a}}\psi^{\prime}_{L}(x),
\nonumber\\
\psi_{R}(x)&=&e^{i\chi^{a}T^{a}}\psi^{\prime}_{R}(x),
\nonumber\\
e^{2i(1/f_{\pi})\pi^{a}(x)T^{a}}&=&
e^{-i\chi^{a}T^{a}}e^{2i(1/f_{\pi}){\pi^{a}(x)}^{\prime}
T^{a}}e^{-i\chi^{a}T^{a}},
\end{eqnarray}
the Lagrangian is form invariant if one sets $e=0$. If one 
imposes this global chiral symmetry, an additional naive mass 
term $m_{0}$ in (3.1) which is obtained by the replacement 
\begin{eqnarray}
mU(\pi) \rightarrow m_{0}+mU(\pi)
\end{eqnarray}
is not allowed. This may be compared to (2.28).
 
We now perform a field-dependent unitary transformation 
\begin{eqnarray}
&&\psi(x)= V(\pi)\psi^{\prime}(x)=V_{R}(\pi)\psi^{\prime}_{R}(x)
+V_{L}(\pi)\psi^{\prime}_{L}(x),\nonumber\\
&&\bar{\psi}(x)=\bar{\psi}^{\prime}(x)V(\pi)
=\bar{\psi}^{\prime}_{R}(x)V_{R}(\pi)^{\dagger}
+\bar{\psi}^{\prime}_{L}(x)V_{L}(\pi)^{\dagger}
\end{eqnarray}
with~\footnote{We have $V_{R}(\pi)=\exp\{-i\frac{1}{f_{\pi}}
\pi^{a}T^{a}\}$ and $V_{L}(\pi)=\exp\{i\frac{1}{f_{\pi}}
\pi^{a}T^{a}\}$ in the fixed chiral frame. If one defines the 
global chiral transformation law by $V_{L}(\pi)\rightarrow 
e^{-i\chi^{a}T^{a}}V_{L}(\pi)$ and $V_{R}(\pi)\rightarrow 
e^{i\chi^{a}T^{a}}V_{R}(\pi)$, the transformation law in (3.6)
is realized if one understands that $\exp\{2i\frac{1}{f_{\pi}}
\pi^{a}T^{a}\}=V_{L}(\pi)V_{R}(\pi)^{\dagger}$ and the fermion 
fields $\psi^{\prime}$ and $\bar{\psi}^{\prime}$ in (3.8) are 
not transformed under the global chiral transformation. }
\begin{eqnarray}
V(\pi)=e^{-i(1/f_{\pi})\gamma_{5}\pi^{a}(x)T^{a}}.
\end{eqnarray}
One then naively obtains the result 
\begin{eqnarray}
&&\int{\cal D}U(\pi) {\cal D}\bar{\psi}{\cal D}\psi
\exp\{\int d^{4}x{\cal L}\}
\nonumber\\
&&=\int{\cal D}U(\pi) {\cal D}\bar{\psi}^{\prime}
{\cal D}\psi^{\prime}
\exp\{\int d^{4}x{\cal L}^{\prime}\}\nonumber\\
&&=\int{\cal D}U(\pi) {\cal D}\bar{\psi}
{\cal D}\psi
\exp\{\int d^{4}x{\cal L}^{\prime}\}
\end{eqnarray}
where
\begin{eqnarray}
{\cal L}^{\prime}&=&
\bar{\psi}(x)[i\gamma^{\mu}(\partial_{\mu}
-ieQA_{\mu}+V^{\dagger}(\pi)D_{\mu}V(\pi))-m]\psi(x)
\nonumber\\
&&+
\frac{f^{2}_{\pi}}{16}{\rm Tr}\partial_{\mu}U(\pi)
\partial^{\mu}U(\pi)^{\dagger}
\end{eqnarray}
with 
\begin{eqnarray}
D_{\mu}V(\pi)=\partial_{\mu}V(\pi)-ie[QA_{\mu},V(\pi)]
\end{eqnarray}
by assuming the invariance of the measure
\begin{eqnarray}
{\cal D}\bar{\psi}{\cal D}\psi
={\cal D}\bar{\psi}^{\prime}{\cal D}\psi^{\prime}.
\end{eqnarray}
We also used the fact that the naming of integral variables 
is arbitarary in the path integral (3.10).

Here we performed a naive manipulation by ignoring the possible
Jacobian for the above change of integration variables (3.8). 
Nevertheless, we obtain the term in (3.11), which was called 
"Dyson term" in the old literature~\cite{fukuda, tomonaga, 
steinberger, schwinger}
\begin{eqnarray}
&&\bar{\psi}(x)i\gamma^{\mu}(
V^{\dagger}(\pi)D_{\mu}V(\pi))\psi(x)
\nonumber\\
&&\sim(1/f_{\pi})\bar{\psi}(x)\gamma^{\mu}\gamma_{5}
(D_{\mu}\pi(x))\psi(x)
\end{eqnarray}
in the order linear in the variables $\pi(x)$ with 
$\pi(x)=\pi^{a}(x)T^{a}$ and 
\begin{eqnarray}
D_{\mu}\pi(x)=\partial_{\mu}\pi(x)-ie[QA_{\mu},\pi(x)].
\end{eqnarray}
The above naive manipulation suggests the equivalence of the 
derivative coupling in ${\cal L}^{\prime}$ (3.11),
\begin{eqnarray}
{\cal L}^{\prime}\sim 
(1/f_{\pi})\bar{\psi}(x)\gamma^{\mu}\gamma_{5}
\partial_{\mu}\pi(x)\psi(x),
\end{eqnarray}
and  the pseudoscalar coupling in the starting Lagrangian 
${\cal L}$ (3.1),
\begin{eqnarray}
{\cal L} \sim -2im(1/f_{\pi})\bar{\psi}(x)\gamma_{5}\pi(x)
\psi(x),
\end{eqnarray}
to the order linear in the pion fields. 

The derivative coupling in ${\cal L}^{\prime}$ (3.11), which 
appears 
sandwiched by fermion fields $\psi^{\dagger}$ and $\psi$, 
precisely corresponds to  the geometric 
term in (2.16), though we here have a four-dimensional 
derivative 
instead of the simple time derivative in (2.16). Naively, the 
appearance of the derivative coupling  is also
 regarded as a result of the failure of the decoupling theorem 
for $m\rightarrow {\rm large}$ in the sense of  Ref.\cite{stone},
but the actual failure of the decoupling theorem
is more involved as will be explained later. 
It is clear that the above operation is a naive one and 
the appearance of the above Dyson term has nothing  to do with 
the quantum anomaly. It is well-known that the above two 
Lagrangians (3.1) and (3.11) give rise to quite different
predictions for the decay amplitude $\pi^{0}\rightarrow \gamma+
\gamma$ in the {\em soft-pion} limit, which marked the genesis 
of the modern notion of quantum anomaly~\cite{bell, adler2}.
This in particular implies that 
\begin{eqnarray}
&&{\rm Det}[i\gamma^{\mu}(\partial_{\mu} -ieQA_{\mu})-mU(\pi)]
\nonumber\\
&&\neq {\rm Det}[i\gamma^{\mu}(\partial_{\mu}
-ieQA_{\mu}+V^{\dagger}(\pi)D_{\mu}V(\pi))-m]
\end{eqnarray}
in contrast to (2.27) and (2.35).

Some of the essential and general properties of the quantum 
anomalies are:\\
1. The anomalies are not recognized by a naive manipulation of 
the classical Lagrangian or action (or by a naive canonical 
manipulation in operator formulation), which leads to the naive 
N\"{o}ther's theorem.\\
2. The quantum anomaly is related to the quantum breaking of 
classical symmetries (and the failure of the naive N\"{o}ther's 
theorem). For example, the Gauss law operator (or BRST charge)
becomes time-dependent and thus it cannot be used to specify 
physical states in anomalous gauge theory~\cite{fujikawa4}.\\
3. The quantum anomalies are generally associated with an 
infinte number of degrees of freedom. The anomalies in the 
practical calculation are thus 
closely related to the regularization, though the anomalies by 
themselves are perfectly finite.\\
4. In the path integral formulation, the anomalies are 
recognized as non-trivial Jacobians for the change of 
path integral variables associated with classical symmetries.

None of these essential properties are shared with the 
derivation of geometric terms in Section 2. Rather, the 
geometric term there (2.16) corresponds to the naive Dyson term 
in (3.11), which is known to fail to account for the whole 
story of the above chiral transformation.
\\

To incorporate the anomaly, one needs to evaluate the Jacobian 
carefully for the above  chiral transformation 
(3.8)~\cite{fujikawa}. 
One may first rewrite the covariant derivative in (3.1) as
\begin{eqnarray}
D_{\mu}=\partial_{\mu}-ieQA_{\mu}=\partial_{\mu}
-ieY A_{\mu}-ieT^{3}A_{\mu}
\end{eqnarray}
with
\begin{eqnarray}
Y=\frac{1}{2}, \ \ \ T^{3}=\frac{1}{2}\sigma^{3}.
\end{eqnarray}
The Wess-Zumino term for the transformation (3.8) then has a 
well-known form~\cite{jackiw, bertlmann, fuji-suzuki} 
\begin{eqnarray}
{\cal D}\bar{\psi}{\cal D}\psi&=&J{\cal D}\bar{\psi}^{\prime}
{\cal D}\psi^{\prime},\nonumber\\
\ln J&=&i\int d^{4}x{\cal L}_{{\rm Wess-Zumino}}\nonumber\\
&=&i\int d^{4}x \int_{0}^{1}ds\frac{1}
{f_{\pi}}\epsilon^{\mu\nu\alpha\beta}{\rm tr}\pi^{a}(x)T^{a}
\frac{1}{16\pi^{2}}
\nonumber\\
&&\times \{
\frac{e^{2}}{2}[U(s)^{\dagger}T^{3}U(s)+U(s)T^{3}U(s)^{\dagger}]
F_{\mu\nu}F_{\alpha\beta}]\nonumber\\
&&\ \ \ \ +4ie[F_{\mu\nu}a_{\alpha}a_{\beta}]\}
\end{eqnarray}
where
\begin{eqnarray}
&&F_{\mu\nu}=\partial_{\mu}A_{\nu}-\partial_{\nu}A_{\mu},
\nonumber\\
&&a_{\alpha}=\frac{i}{2}[U(s)^{\dagger}D_{\alpha}U(s)
-U(s)(D_{\alpha}U(s))^{\dagger}]
\end{eqnarray}
with 
\begin{eqnarray}
&&U(s)\equiv e^{-is(1/f_{\pi})\pi^{a}(x)T^{a}},\nonumber\\
&&D_{\alpha}U(s)=\partial_{\alpha}U(s)-ie[A_{\alpha}T^{3}, U(s)].
\end{eqnarray}
This is obtained by an integral of the Jacobian for the repeated 
applications of the infinitesimal 
transformation
\begin{eqnarray}
\psi(x)&=& e^{-ids (1/f_{\pi})\pi^{a}(x)T^{a}\gamma_{5}}
\psi^{\prime}(x),\nonumber\\
\bar{\psi}(x)&=&\bar{\psi}^{\prime}(x) 
e^{-ids (1/f_{\pi})\pi^{a}(x)T^{a}\gamma_{5}},
\end{eqnarray}
and ${\rm tr}$  stands for the trace over the $2\times2$ 
matrices with ${\rm tr}T^{a}T^{b}=\frac{1}{2}\delta_{ab}$.

In terms of the functional determinant we have~\footnote{In the 
present chiral
$SU(2)$ symmetry, which is anomaly free by itself, no 
Wess-Zumino term arises for $A_{\mu}=0$. For $SU(3)$, for 
example, one obtains a 
non-trivial Jacobian or the Wess-Zumino term even with 
$A_{\mu}=0$, and such a term is shown to exhibit a topological 
property~\cite{jackiw, bertlmann}.}
\begin{eqnarray}
&&{\rm Det}[i\gamma^{\mu}(\partial_{\mu} -ieQA_{\mu})-mU(\pi)]
\nonumber\\
&&= {\rm Det}[i\gamma^{\mu}(\partial_{\mu}
-ieQA_{\mu}+V^{\dagger}(\pi)D_{\mu}V(\pi))-m]\nonumber\\
&&\ \ \ \times\exp\{i\int d^{4}x{\cal L}_{\rm Wess-Zumino}\}
\end{eqnarray} 
which may be compared to (2.27). For $T\rightarrow \rm large$
(in the present Euclidean theory), we have 
\begin{eqnarray}
&&{\rm Det}[i\gamma^{\mu}(\partial_{\mu}
-ieQA_{\mu}+V^{\dagger}(\pi)D_{\mu}V(\pi))-m]\nonumber\\
&&\ \ \ \times\exp\{i\int d^{4}x{\cal L}_{\rm Wess-Zumino}\}
\nonumber\\
&&\sim \exp\{-E_{\rm vac}T\}
\exp\{i\int d^{4}x{\cal L}_{\rm Wess-Zumino}\}\nonumber\\
&&=\exp\{i\int d^{4}x{\cal L}_{\rm Wess-Zumino}\}
\end{eqnarray}
for a fixed large $m$ and slowly varying $\pi(x)$ with 
periodic boundary conditions, for which
we have a mass gap $\sim m$ and thus the fermionic vacuum with
vanishing energy is isolated. This relation may be compared to 
(2.38).

In the order linear in the pion fields, we have the Jacobian
\begin{eqnarray}
\ln J&=&i\int d^{4}x \frac{1}{f_{\pi}}({\rm tr}
T^{a}T^{3})\pi^{a}(x)
\frac{e^{2}}{16\pi^{2}}
\epsilon^{\mu\nu\alpha\beta}F_{\mu\nu}F_{\alpha\beta}
\nonumber\\
&=&i\int d^{4}x \frac{1}{f_{\pi}}\pi^{0}(x)
\frac{e^{2}}{32\pi^{2}}
\epsilon^{\mu\nu\alpha\beta}F_{\mu\nu}F_{\alpha\beta}.
\end{eqnarray}
It is well known that this Wess-Zumino term (3.27) when added to
 the Lagrangian ${\cal L}^{\prime}$ in (3.11)
\begin{eqnarray}
&&\int d^{4}x [{\cal L}^{\prime}+{\cal L}_{\rm Wess-Zumino}]
\nonumber\\
&\sim& 
-\int d^{4}x 
(1/f_{\pi})\pi^{0}(x)\partial_{\mu}[\bar{\psi}(x)\gamma^{\mu}
\gamma_{5}T^{3}\psi(x)] \nonumber\\
&&+ \int d^{4}x \frac{i}{f_{\pi}}\pi^{0}(x)
\frac{e^{2}}{32\pi^{2}}
\epsilon^{\mu\nu\alpha\beta}F_{\mu\nu}F_{\alpha\beta} 
\end{eqnarray}  
correctly describes the decay 
$\pi^{0}\rightarrow \gamma + \gamma$ in the soft-pion 
limit~\cite{bell, adler2} in agreement with the result on 
the basis of (3.1). 
In the operator notation, this equivalence is expressed as
the relation
\begin{eqnarray}
\partial_{\mu}[\bar{\psi}(x)\gamma^{\mu}
\gamma_{5}T^{3}\psi(x)]=2im[\bar{\psi}(x)\gamma_{5}T^{3}\psi(x)]
+i\frac{e^{2}}{32\pi^{2}}
\epsilon^{\mu\nu\alpha\beta}F_{\mu\nu}F_{\alpha\beta}
\end{eqnarray}
which expresses the failure of the naive N\"{o}ther's theorem for
the exact global chiral symmetry generated by $\gamma_{5}T^{3}$.

It is instructive to see  in detail how this 
equivalence in the 
decay $\pi^{0}\rightarrow \gamma + \gamma$ is 
achieved. We consider two distinct cases:\\
(i)$m \neq 0$\\
In this case, the operator $[\bar{\psi}(x)\gamma^{\mu}
\gamma_{5}T^{3}\psi(x)]$ in (3.28) is free of infrared 
singularity in the soft-pion limit, namely, for the 
four-momentum of the pion $p_{\mu}\sim 0$. Thus the 
first term in (3.28) vanishes in the soft-pion limit 
\begin{eqnarray}
\lim_{p_{\mu}\rightarrow 0}\int d^{4}x e^{ip_{\mu}x^{\mu}}
\partial_{\mu}[\bar{\psi}(x)\gamma^{\mu}
\gamma_{5}T^{3}\psi(x)] =0
\end{eqnarray}
and the second anomaly term gives the same result as the 
pseudo-scalar coupling in (3.17).\\
(ii) $m=0$ \\
This case is singular from the point of view of spontaneously 
broken chiral symmetry. Nevertheless, in this case, which 
corresponds to the case $\mu=0$ in (2.2),
the pseudo-scalar coupling in (3.17) vanishes. On the other 
hand, the current $[\bar{\psi}(x)\gamma^{\mu}
\gamma_{5}T^{3}\psi(x)]$ becomes singular in the soft-pion 
limit but still  one can use the operator 
relation (3.29) with $m=0$ in (3.28) and the two terms in (3.28)
 cancel each other, to be consistent with the vanishing 
pseudo-scalar coupling in (3.17). 

We emphasize that the derivation of the chiral anomaly does not
depend on the relative magnitude of $m$, which is analogous to
$\mu$ in (2.2), and the frequency of the external variables
such as the gauge field, but rather depends on the relative 
magnitude of the cut-off mass $M$, such as the Pauli-Villars 
regulator mass which can be chosen to be arbitrarily large, and 
the frequency of the external variables. This is quite different 
from the case of the geometric phase where the parameter $\mu$, 
which corresponds to $m$, directly enters the criterion of the 
adiabatic approximation in (2.20) where $1/T$ corresponds to
the frequency of the external variables. Because of this 
difference, the {\em failure} of the  decoupling theorem in the 
chiral 
anomaly is stated precisely as follows: The derivative coupling 
term in (3.28), which corresponds to the geometric term in 
(2.16), vanishes for $m\rightarrow\infty$~\cite{adler}, and the 
anomaly term balances the pseudo-scalar term in (3.17) which  
does not vanish in the limit. 
   
The discovery of the chiral anomaly is based on the 
recognition that the naive Dyson's relation, namely the naive
equivalence between (3.1) and (3.11), inevitably fails in 
gauge field theory, and one needs to include an extra 
Jacobian (or Wess-Zumino term).

\section{Discussion}

We have explained the basic differences between the 
geometric phase and the quantum anomaly by analyzing the 
concrete model due to Stone and a corresponding model in field 
theory which contains a true quantum anomaly. The only 
similarity between the geometric phase in 
the adiabatic approximation and the Wess-Zumino term is that 
both of them exhibit topological properties under certain 
limiting conditions.

In contrast, the differences are more profound and fundamental.
Firstly, the geometric phase arises from the {\em naive} 
rearrangement of terms inside the fermionic 
operator sandwiched by $\psi^{\dagger}$ and $\psi$, whereas the 
Wess-Zumino term associated with quantum anomaly 
arises from the Jacobian, namely, a completely new additional 
part to the Lagrangian. Secondly, the geometric phase is 
recognized only 
when one throws away a part of the original Hamiltonian in the 
adiabatic approximation, whereas the quantum anomaly is exact 
without any approximation. The Born-Oppenheimer approximation in
the geometric phase means a neglect of a part of the 
Hamiltonian, 
whereas the Born-Oppenheimer approximation in the quantum 
anomaly, if any~\cite{nelson}, is actually not an approximation;
 this is obvious in the path integral formulation of quantum 
anomalies where the Born-Oppenheimer approximation simply means 
a specific order of path integration, namely, one first 
integrates over fermions with fixed bosonic (such as gauge or 
Nambu-Goldstone) variables and then one integrates over bosonic 
variables later. The path integral over the fermionic variables,
 which are quadratic, can be performed exactly and the 
Wess-Zumino term is induced by this fermionic path integral; in 
this sense  no approximation is invloved in the analysis of the 
chiral anomaly, though the path integral of bosonic variables in
 the non-linear effective chiral model in Section 3 is not 
renormalizable. Because of this difference, the topological 
property of the geometric phase is inevitably trivial in the 
practical Born-Oppenheimer approximation for any finite time 
interval $T$ if one deals with the exact Hamiltonian, whereas 
the topology in the quantum anomaly, which is basically a short 
distance phenomenon in four-dimesional space-time, is exact once
its existence is established since no approximation is involved. 

One may still wonder, if our assertion is valid, what then 
happens with the analysis by Nelson and 
Alvarez-Gaume~\cite{nelson} where a 
precise analogy between the quantum anomaly in the Hamiltonian 
interpretation and the geometric 
phase is forcefully argued. We believe that all what are 
said about the chiral anomaly there~\cite{nelson} are correct. 
We also believe that they can perform all of their analyses of 
the chiral anomaly without referring to the geometric phase in 
quantum mechanics. The pair production picture in~\cite{nelson} 
is based on the fact that one can arrive at the level crossing 
point with vanishing energy at a fixed well-defined time 
$t_{0}$. This means the {\em failure} of the naive 
adiabatic picture as is emphasized in~\cite{nelson}. On the 
other hand, the validity of the 
topological property of the geometric phase is based on the 
condition that we never approach the level crossing point for 
any finite $t$, namely, on the strict validity of the adiabatic 
picture. The topological property of the geometric phase cannot 
be used in the context of the analysis in~\cite{nelson}. If one 
arrives at the level crossing at finte 
$t=t_{0}$, for example, one can suitably re-define the time 
variable and smoothly deform the background variable such that 
the period $T$ of the background variable  is finite. This is 
generally the case in the mathematical analysis of the index 
theorem~\cite{atiyah} which is based on the compact Euclidean 
space-time such as $S^{4}$. For a {finite} time interval $T$, 
the topological property of the geometric phase, such as the 
topological proof~\cite{stone2} of the Longuet-Higgins phase
 change rule, fails as we have shown elsewhere~\cite{fujikawa2} 
and also in Section 2 of the present paper. The topological 
property of the geometric phase crucially depends on the 
very precise definition of the adiabatic approximation; the 
movement of the external parameter must be infinitely slow, 
{\em i.e.}, not only the period 
$T\rightarrow\infty$ but also the variation of the 
background variable at each moment is negligible~\cite{simon}. 

As is clear in the analysis in~\cite{nelson},
the quantum anomaly influences all the states of the Fock space 
equally, whereas the geometric phase appears only in the 
specific states of the Fock space and does not influence the 
vacuum state. The form of the geometric phase is also 
state-dependent. See eq.(2.38). Also, the geometric phase is 
indepedent of the parameter $\mu_{0}$ as in (2.33), whereas the 
analysis of quantum anomaly in~\cite{nelson} crucially depends 
on the crossing of vanishing eigenvalues in chiral gauge
theory which corresponds to the specific choice $\mu_{0}=0$ in 
the context of the model in Section 2. The general level 
crossing problem in the context of geometric phases is regarded 
to be related to the vector-like transformation as is seen in 
(2.33) and (2.35). See also (A.9). In contrast the chiral 
symmetry, not the vector-like symmetry, is crucial in the 
analysis of the anomaly.
 The level crossing by itself has no connection to the anomaly.
  
We also note that all 
the properties of the chiral anomaly are understood in terms of 
the Green's functions instead of going to the S-matrix. The 
Green's functions are the statements about the local properties 
of field theory unlike the S-matrix which involves a subtle limit
of the infinite time interval in field theory. The global 
$SU(2)$ anomaly by Witten~\cite{witten} may appear to depend on 
the infinite time interval to some extent, but  
we note that the global $SU(2)$ anomaly is also known to 
be described by the Wess-Zumino term related to the group 
$SU(3)$, which is defined in the framework of Green's functions, 
in a suitable fomulation of the problem~\cite{witten2,elitzur}. 
The quantum anomaly, as we understand it in gauge field theory, 
is a precise statement and as such it should not depend on the 
technical details of the adiabatic approximation, unlike the 
case of the geometric phase associated with level crossing in 
quantum mechanics. 

There are well-known odd-dimensional cousins of chiral anomalies,
namely, the Chern-Simons terms which exhibit topological 
properties. The Chern-Simons terms induced by fermions, which 
are sometimes called parity anomaly, or added by hand are 
closely related to the chiral anomaly not only by the descent 
formula~\cite{jackiw, bertlmann} or dimensional reduction but 
also in the explicit Feynman diagramatic calculations. If one 
provides a precise
definition of the geometric (or Berry) phase in general field 
theoretical contexts, possibly asking some association with 
level crossing and adiabaticity as minimal requirements, it 
would be possible to analyze  the relation between the Berry 
phase and the odd-dimensional cousins of chiral anomaly.

As an explicit example of the geometric phase in realistic
condensed matter physics, we mention the recent works on 
anomalous Hall effect~\cite{nagaosa}. In those works, readers 
will find that all the basic ingredients of the geometric 
(or Berry) phase, such as level crossing, adiabaticity and 
approximate topology, are contained. This class of models are 
included in the general model in Appendix of the present paper, 
and thus our analysis in the present paper is applicable to them.
 
Finally, we note that  the Aharonov-Bohm phase is 
topologically exact even for a finite time interval $T$ unlike
the geometric phase.  The 
Aharonov-Bohm effect contains an extra dynamical
freedom, namely, the electromagnetic potential which is 
{\em time-independent}, and the space for the Aharonov-Bohm 
effect is not simply connected. None of these crucial  
features are shared with the geometric phase, though certain 
feature of the Aharonov-Bohm effect is known to be shared with 
the geometric phase~\cite{berry}. We think that a clear 
distinction between the Aharonov-Bohm phase and the geometric 
phase is also important, since the notion of winding number is 
defined for the Aharonov-Bohm phase whereas
no notion of winding number in the geometric phase for
any finite time interval $T$ as the topology is trivial.

\section{Conclusion}

The model in Ref.~\cite{stone}, which is essentially identical 
to the simplest example discussed by Berry in his original 
paper~\cite{berry}, shows that the Berry phase associated with 
level crossing gives the topological phase for certain states in
 the Fock space in the precise adiabatic limit. The phase factor
 has the same form as the anomaly-induced Wess-Zumino term 
appearing in certain field theoretical models. The key concepts 
involved in the model, namely, the level crossing, topology and 
adiabatic approximation also appear in the Hamiltonian analysis 
of chiral anomalies by Nelson and Alvarez-Gaume~\cite{nelson}. 
This fact led to an expectation that the very basic mechanism of 
chiral anomalies, which have been established by the efforts of 
various authors, notably by Bell and Jackiw~\cite{bell} and 
Adler~\cite{adler2}, may be identified with the basic mechanism 
of the adiabatic Berry phase related to level crossing in the 
simple Schr\"{o}dinger problem. What we have shown in the 
present paper is that this expectation is not realized, and the 
similarity between the two is superficial. We
have first explained the difference between the two on the 
basis of general characteristics of chiral anomaly, such as the 
failure of the naive manipulation and the failure of the 
decoupling theorem, by using two explicit examples
 in Sections 2 and 3.  Our conclusion is valid for a more 
general class of level crossing problems summarized in Appendix.
We then explained the difference between the two from the point
of view of level crossing and topology.
The difference between the chiral anomaly 
and the Berry phase is simply stated as follows: The 
topology of gauge fields leads to level crossing in the 
fermionic sector in the case of chiral anomaly and the 
{\em failure} of the adiabatic approximation is essential in the 
analysis, whereas the (potential) level crossing in the matter 
sector leads to the topology of the Berry phase only when the 
very precise adiabatic approximation holds. These two cannot be 
compatible with each other.

In the early literature on the geometric phase, the similarity 
between the geometric phase and the quantum anomaly, though  
rather superficial one, was emphasized~\cite{jackiw2}. That 
analogy was useful at the initial developing stage of the 
subject. But in view of the wide use of the 
terminology ``geometric phase'' in various fields in physics 
today~\cite{review}, it is our opinion that a more precise 
distinction of various loosely related phenomena is also 
desirable. To be precise, what we are suggesting is to call
chiral anomaly as chiral anomaly,
Wess-Zumino term as Wess-Zumino term, Chern-Simons term as 
Chern-Simons term, and Aharovov-Bohm phase as Aharonov-Bohm
phase, etc., since those terminologies convey  very clear 
messages and well-defined physical contents which the majority 
in physics community can readily recognize. Even in this sharp 
definition of terminology, one can still clearly identify the 
geometric (or Berry) phase and its physical characteristics, 
which cannot be described by other notions, as 
the concrete physical example in Ref.~\cite{nagaosa} suggests.

We believe that a sharp definition of the scientific term 
``geometric phase'', probably by asking some association with 
level crossing and adiabaticity as minimal requirements, is also
 important for those experts working on the geometric phase 
itself, since then the wider audience can easily 
identify the phenomena, which are intrinsic to the geometric 
phase and  cannot be described by other notions, and  
consequently they will appreciate more the usefulness of the 
geometric phase.

\appendix

\section{General Level Crossing Problem}

The general geometric phase associated with 
any level crossing in the second quantized formulation exhibits 
the same topological properties as the specific example in 
Section 2; 
approximate monopole-like behavior in the adiabatic 
approximation but 
actually topologically trivial in the infinitesimal neighborhood
of level crossing for any finite time interval 
$T$. This property may be relevant to the analysis in 
Ref.\cite{aitchison}, where the geometric phase is used as an 
{\em analogue} of the Wess-Zumino term, and we sketch the 
analysis of the general level 
crossing~\cite{fujikawa2, fujikawa3} in this appendix:  

We start with the generic hermitian Hamiltonian 
\begin{equation}
\hat{H}=\hat{H}(\hat{\vec{p}},\hat{\vec{x}},X(t))
\end{equation}
for a single particle theory in the  background 
variable $X(t)=(X_{1}(t),X_{2}(t),...)$.
The path integral for this theory for the time interval
$0\leq t\leq T$ in the second quantized 
formulation is given by 
\begin{eqnarray}
Z&=&\int{\cal D}\psi^{\star}{\cal D}\psi
\exp\{\frac{i}{\hbar}\int_{0}^{T}dtd^{3}x[
\psi^{\star}(t,\vec{x})i\hbar\frac{\partial}{\partial t}
\psi(t,\vec{x})\nonumber\\
&&-\psi^{\star}(t,\vec{x})
\hat{H}(\frac{\hbar}{i}\frac{\partial}{\partial\vec{x}},
\vec{x},X(t))\psi(t,\vec{x})] \}.
\end{eqnarray}
We then define a complete set of eigenfunctions
\begin{eqnarray}
&&\hat{H}(\frac{\hbar}{i}\frac{\partial}{\partial\vec{x}},
\vec{x},X(0))u_{n}(\vec{x},X(0))
=\lambda_{n}u_{n}(\vec{x},X(0)), \nonumber\\
&&\int d^{3}xu_{n}^{\star}(\vec{x},X(0))u_{m}(\vec{x},X(0))=
\delta_{nm},
\end{eqnarray}
and expand 
\begin{eqnarray}
\psi(t,\vec{x})=\sum_{n}a_{n}(t)u_{n}(\vec{x},X(0)).
\end{eqnarray}
We then have
\begin{eqnarray}
{\cal D}\psi^{\star}{\cal D}\psi=\prod_{n}{\cal D}a_{n}^{\star}
{\cal D}a_{n}
\end{eqnarray}
and the path integral is written as 
\begin{eqnarray}
Z&=&\int \prod_{n}{\cal D}a_{n}^{\star}
{\cal D}a_{n}
\exp\{\frac{i}{\hbar}\int_{0}^{T}dt[
\sum_{n}a_{n}^{\star}(t)i\hbar\frac{\partial}{\partial t}
a_{n}(t)\nonumber\\
&&-\sum_{n,m}a_{n}^{\star}(t)E_{nm}(X(t))a_{m}(t)] \}
\end{eqnarray}
where 
\begin{eqnarray}
E_{nm}(X(t))=\int d^{3}x u_{n}^{\star}(\vec{x},X(0))
\hat{H}(\frac{\hbar}{i}\frac{\partial}{\partial\vec{x}},
\vec{x},X(t))u_{m}(\vec{x},X(0)).
\end{eqnarray}

We next perform a unitary transformation
\begin{eqnarray}
a_{n}=\sum_{m}U(X(t))_{nm}b_{m}
\end{eqnarray}
where 
\begin{eqnarray}
U(X(t))_{nm}=\int d^{3}x u^{\star}_{n}(\vec{x},X(0))
v_{m}(\vec{x},X(t))
\end{eqnarray}
with the instantaneous eigenfunctions of the Hamiltonian
\begin{eqnarray}
&&\hat{H}(\frac{\hbar}{i}\frac{\partial}{\partial\vec{x}},
\vec{x},X(t))v_{n}(\vec{x},X(t))
={\cal E}_{n}(X(t))v_{n}(\vec{x},X(t)), \nonumber\\
&&\int d^{3}x v^{\star}_{n}(\vec{x},X(t))v_{m}(\vec{x},X(t))
=\delta_{n,m}.
\end{eqnarray}
We can thus re-write the path integral as 
\begin{eqnarray}
&&Z=\int \prod_{n}{\cal D}b_{n}^{\star}{\cal D}b_{n}
\exp\{\frac{i}{\hbar}\int_{0}^{T}dt[
\sum_{n}b_{n}^{\star}(t)i\hbar\frac{\partial}{\partial t}
b_{n}(t)\nonumber\\
&&+\sum_{n,m}b_{n}^{\star}(t)
\langle n|i\hbar\frac{\partial}{\partial t}|m\rangle
b_{m}(t)-\sum_{n}b_{n}^{\star}(t){\cal E}_{n}(X(t))b_{n}(t)] \}
\end{eqnarray}
where the second term in the action, which is defined by 
\begin{eqnarray}
\int d^{3}x v^{\star}_{n}(\vec{x},X(t))
i\hbar\frac{\partial}{\partial t}v_{m}(\vec{x},X(t))
\equiv \langle n|i\hbar\frac{\partial}{\partial t}|m\rangle,
\end{eqnarray}
 stands for the geometric term. We take the time $T$ as a period
 of the  variable $X(t)$. The adiabatic 
process means that $T$ is much larger than the typical
time scale $\hbar/\Delta{\cal E}_{n}(X(t))$.
The result (A.11) is also directly obtained by the expansion 
\begin{eqnarray}
\psi(t,\vec{x})=\sum_{n}b_{n}(t)v_{n}(\vec{x},X(t)).
\end{eqnarray} 

In the operator formulation,
we thus obtain the effective Hamiltonian (depending on Bose or 
Fermi statistics)
\begin{eqnarray}
\hat{H}_{eff}(t)&=&\sum_{n}\hat{b}_{n}^{\dagger}(t)
{\cal E}_{n}(X(t))\hat{b}_{n}(t)\nonumber\\
&&-\sum_{n,m}\hat{b}_{n}^{\dagger}(t)
\langle n|i\hbar\frac{\partial}{\partial t}|m\rangle
\hat{b}_{m}(t)
\end{eqnarray}
with $[\hat{b}_{n}(t), \hat{b}^{\dagger}_{m}(t)]_{\mp}
=\delta_{n,m}$.
All the information about geometric phases  is included in 
the effective Hamiltonian and thus geometric phases are purely {\em dynamical}.

When one defines the Schr\"{o}dinger picture 
$\hat{{\cal H}}_{eff}(t)$ by replacing all $\hat{b}_{n}(t)$ by
$\hat{b}_{n}(0)$ in $\hat{H}_{eff}(t)$,
the second quantization formula for the evolution operator 
gives rise to~\cite{fujikawa2, fujikawa3}  
\begin{eqnarray}
&&\langle m|T^{\star}\exp\{-\frac{i}{\hbar}\int_{0}^{t}
\hat{{\cal H}}_{eff}(t)
dt\}|n\rangle\nonumber\\ 
&&=
\langle m(t)|T^{\star}\exp\{-\frac{i}{\hbar}\int_{0}^{t}
\hat{H}(\hat{\vec{p}}, \hat{\vec{x}},  
X(t))dt \}|n(0)\rangle 
\end{eqnarray}
where $T^{\star}$ stands for the time ordering operation, and 
the state vectors in the second quantization  on the left-hand 
side are defined by $
|n\rangle=\hat{b}_{n}^{\dagger}(0)|0\rangle$,
and the state vectors on the right-hand side  stand for the 
first quantized states defined by
$\langle\vec{x}|n(t)\rangle=v_{n}(\vec{x},(X(t))$.
Both-hand sides of the above equality (A.15) are exact, but the 
difference is that the geometric term, both of diagonal and 
off-diagonal, is explicit in the second quantized formulation 
on the left-hand side.

The probability amplitude which satisfies Schr\"{o}dinger 
equation is given by
\begin{eqnarray}
\psi_{n}(\vec{x},t; X(t))=
\langle 0|\hat{\psi}(t,\vec{x})\hat{b}^{\dagger}_{n}(0)|0\rangle
\end{eqnarray}
since $i\hbar\partial_{t}\hat{\psi}=\hat{H}\hat{\psi}$ in the 
present problem.
In the adiabatic approximation,
where we assume the dominance of diagonal elements, we have 
(see also~\cite{kuratsuji}) 
\begin{eqnarray}
&&\psi_{n}(\vec{x},t; X(t))\nonumber\\
&&=\sum_{m} v_{m}(\vec{x};X(t))
\langle m(t)|T^{\star}\exp\{-\frac{i}{\hbar}\int_{0}^{t}
\hat{H}(\hat{\vec{p}}, \hat{\vec{x}},  
X(t))dt \}|n(0)\rangle\nonumber\\
&&=\sum_{m} v_{m}(\vec{x};X(t))
\langle m|T^{\star}\exp\{-\frac{i}{\hbar}\int_{0}^{t}
\hat{{\cal H}}_{eff}(t)dt\}|n\rangle
\nonumber\\
&&\simeq v_{n}(\vec{x};X(t))
\exp\{-\frac{i}{\hbar}\int_{0}^{t}[{\cal E}_{n}(X(t))
-\langle n|i\hbar\frac{\partial}{\partial t}|n\rangle]dt\}.
\end{eqnarray}
by noting (A.15).

The path integral formula (A.11) is based on the expansion
 (A.13) and the starting path integral (A.2) depends only on 
the field
variable $\psi(t,\vec{x})$, not on  $\{ b_{n}(t)\}$
and $\{v_{n}(\vec{x},X(t))\}$ separately. This fact shows that 
our formulation contains an exact hidden local gauge symmetry 
\begin{eqnarray}
&&v_{n}(\vec{x},X(t))\rightarrow v^{\prime}_{n}(t; \vec{x},X(t))=
e^{i\alpha_{n}(t)}v_{n}(\vec{x},X(t)),\nonumber\\
&&b_{n}(t) \rightarrow b^{\prime}_{n}(t)=
e^{-i\alpha_{n}(t)}b_{n}(t), \ \ \ \ n=1,2,3,...,
\end{eqnarray}
where the gauge parameter $\alpha_{n}(t)$ is a general 
function of $t$.  One can confirm that the action 
and the path integral measure in (A.11) are both invariant under
 this gauge transformation. 
 This local symmetry is exact as long as the 
basis set is not singular, and thus it is 
particularly useful in the general adiabatic approximation 
defined by the condition that the basis set (A.10) is 
well-defined~\footnote{This symmetry is a statement that the choice of the coordinates in the functional space is arbitrary in field theory. This symmetry by itself does not imply any conservation law. If one neglects the off-diagonal parts of the geometric term, the theory becomes invariant under $b_{n}(t) \rightarrow 
b^{\prime}_{n}=e^{-i\alpha_{n}}b_{n}(t)$ for a constant 
$\alpha_{n}$ with fixed $v_{n}(\vec{x},X(t))$, and then the symmetry implies a (rather trivial) conservation law, namely, no level crossing.}. The specific basis set (A.10) becomes singular on 
top of level crossing. Of course, one may  
consider a new hidden local gauge symmetry when one defines a new
regular basis set in the neighborhood of the singularity, and 
the freedom in the phase choice of the new basis set persists.
  
The above hidden local gauge symmetry (A.18) is an exact 
symmetry of quantum theory, and thus physical observables in the
 adiabatic approximation should respect this symmetry. Also, by 
using this local gauge freedom, 
one can choose the phase convention of the basis set 
$\{v_{n}(t,\vec{x},X(t))\}$ at one's will such that the analysis
 of geometric phases becomes simplest. 

Our next observation is that $\psi_{n}(\vec{x},t; X(t))$  
transforms under the hidden local gauge symmetry (A.18) as
\begin{eqnarray} 
\psi^{\prime}_{n}(\vec{x},t; X(t))=e^{i\alpha_{n}(0)}
\psi_{n}(\vec{x},t; X(t))
\end{eqnarray}
{\em independently} of the value of $t$. 
This transformation is derived by using the exact 
representation (A.16). This transformation is explicitly checked
 for the adiabatic approximation (A.17) also.

Thus the product
\begin{eqnarray}
\psi_{n}(\vec{x},0; X(0))^{\star}\psi_{n}(\vec{x},T; X(T))
\end{eqnarray}
defines a manifestly gauge invariant quantity, namely, it is 
independent of the choice of the phase convention of the 
complete basis set $\{v_{n}(t,\vec{x},X(t))\}$. One may employ 
this (rather strong) gauge invariance condition as the basis of 
the analysis  of geometric phases, which is shown to replace the 
notions of parallel transport and holonomy~\cite{fujikawa3}. 
Our hidden local gauge symmetry is a symmetry 
of quantum theory and that the Schr\"{o}dinger amplitude 
$\psi_{n}(\vec{x},t; X(t))$ stays in the same 
ray under an arbitrary hidden local gauge transformation of the 
basis set as is shown in (A.19).

For the adiabatic formula (A.17), the gauge invariant quantity 
(A.20) is given by
\begin{eqnarray}
&&\psi_{n}(\vec{x},0; X(0))^{\star}\psi_{n}(\vec{x},T; X(T))
\nonumber\\
&&=v_{n}(0,\vec{x}; X(0))^{\star}v_{n}(T,\vec{x};X(T))
\nonumber\\
&&\times\exp\{-\frac{i}{\hbar}\int_{0}^{T}[{\cal E}_{n}(X(t))
-\langle n|i\hbar\frac{\partial}{\partial t}|n\rangle]dt\}
\end{eqnarray}
where we used the notation $v_{n}(t,\vec{x};X(t))$ to emphasize 
the use of arbitrary gauge in this expression.
We then observe that by choosing the gauge such that 
$v_{n}(T,\vec{x};X(T))=v_{n}(0,\vec{x}; X(0))$
the prefactor 
$v_{n}(0,\vec{x}; X(0))^{\star}v_{n}(T,\vec{x};X(T))$ becomes 
real and positive. Note that we are assuming the cyclic 
evolution of
the external parameter, $X(T)=X(0)$. Then the factor
\begin{eqnarray}
\exp\{-\frac{i}{\hbar}\int_{0}^{T}[{\cal E}_{n}(X(t))
-\langle n|i\hbar\frac{\partial}{\partial t}|n\rangle]dt\}
\end{eqnarray}
extracts all the information about the phase in (A.21) and 
defines a physical quantity. After this gauge fixing, the 
above quantity (A.22) is still invariant under residual gauge 
transformations satisfying the periodic boundary condition
$\alpha_{n}(0)=\alpha_{n}(T)$,
in particular, for a class of gauge transformations defined 
by $\alpha_{n}(X(t))$. Note that our gauge transformation in 
(A.18), which is defined by an arbitrary function 
$\alpha_{n}(t)$,  is much more general. 

In the analysis of  the behavior in the infinitesimal 
neighborhood of a 
specific level crossing, one may truncate the above general
model to a two-level model containing the two levels at 
issue, and  the present formulation (A.14) is essentially 
reduced to the model (2.17) or (2.18); one then finds the 
same approximate topological property for any finite $T$ as in 
the model (2.17). This is explained in detail in 
Ref.~\cite{fujikawa2}.

Based on the above general analysis, the essence of geometric 
phase may be summarized as follows: One obtains an interesting
universal view such as in (A.22) about various level crossing 
problems by making an {\em approximation} (adiabatic 
approximation), which is not clearly seen in the exact 
treatment on the right-hand side of (A.15).

\end{document}